\documentclass[aps,prl,twocolumn,superscriptaddress]{revtex4}
\pdfoutput=1
\usepackage{amsfonts,amssymb,epsfig,amsmath,graphics,subfig}
\usepackage{color}
\usepackage{hyperref,graphicx}
\usepackage{cancel}

\renewcommand{\text}[1]{#1}

\newcommand{\be}{\begin{equation}}
\newcommand{\ee}{\end{equation}}
\newcommand{\ben}{\begin{displaymath}}
\newcommand{\een}{\end{displaymath}}
\newcommand{\bea}{\begin{eqnarray}}
\newcommand{\eea}{\end{eqnarray}}
\newcommand{\bean}{\begin{eqnarray*}}
\newcommand{\eean}{\end{eqnarray*}}
\newcommand{\nn}{\nonumber \\}
\newcommand{\ba}{\begin{array}}
\newcommand{\ea}{\end{array}}
\newcommand{\bi}{\begin{itemize}}
\newcommand{\ei}{\end{itemize}}

\begin{document}
\title{Helical superconducting black holes}

\author{Aristomenis Donos}
\affiliation{Theoretical Physics Group, Blackett Laboratory,
 Imperial College, London SW7 2AZ, U.K.}
  \author{Jerome P. Gauntlett}
\affiliation{Theoretical Physics Group, Blackett Laboratory,
  Imperial College, London SW7 2AZ, U.K.}
\begin{abstract}
We construct novel static, asymptotically $AdS_5$ black hole solutions with Bianchi VII$_0$ symmetry 
that are holographically dual to superconducting phases in four spacetime dimensions with a helical $p$-wave order. 
We calculate the precise temperature dependence of the pitch of the helical order. At zero temperature the black holes have vanishing entropy and approach domain wall solutions that reveal homogenous, non-isotropic dual ground states with emergent scaling symmetry.
\end{abstract}
\maketitle

\setcounter{equation}{0}


\section{Introduction}
The AdS/CFT correspondence is a powerful tool to analyse strongly coupled quantum field
theories and there has been a surge of activity aimed at finding possible applications
both to condensed matter systems and to QCD. One focus has been to holographically realise
various kinds of phases via the construction of fascinating new classes of 
black hole solutions, which are also of interest in their own right. 

An important development was the discovery of black brane solutions that are holographically dual
to superconducting phases, or more precisely, superfluid phases \cite{Gubser:2008px}.
These black holes carry a halo of charged hair which
spontaneously breaks a global abelian symmetry of the dual field theory.
In the simplest examples, the charged hair is in the guise of a bulk scalar field, corresponding to a scalar order parameter
in the dual field theory, and hence an s-wave superconducting phase. $p$-wave superconducting phases, in which the order parameter has angular momentum $l=1$, 
have also been realised \cite{Gubser:2008zu,Aprile:2010ge}.

Spatially modulated phases are also widely seen in Nature. The order parameters for these phases are associated with non-zero momentum and
spontaneously break some or all of the translation invariance. Common examples in condensed matter include spin density waves and charge density waves, 
while QCD at high baryonic density is anticiapted to be in a chiral-wave state \cite{Deryagin:1992rw}.
Spatially modulated phases that are also superconducting are possible \cite{Fulde:1964zz}
and such FFLO phases have been argued to be realised in several systems \cite{fflorev}.
Of particular interest here are $p$-wave superconducting phases with a helical order. In these phases the $l=1$ order parameter points in a given direction in a plane which then rotates as one moves along the direction orthogonal to the plane.
They have been discussed, for example, in the context of non-centrosymmetric heavy fermion compounds \cite{helsc}.

Holographic studies of spatially modulated phases were initiated in \cite{Domokos:2007kt}.
The purpose of this letter is to present the very first construction of fully back-reacted black hole solutions that are holographically dual to spatially modulated phases which are, moreover, superconducting. 
We will consider a class of gravitational models in $D=5$ that couple a metric with a gauge-field and a two-form potential, which have been shown, using a linearised analysis, 
to admit black brane solutions which are dual to $p$-wave superfluid phases with a helical order in $d=4$ \cite{Donos:2011ff}.

Our construction of the new black holes allows us to show that the helical $p$-wave superconducting phase is thermodynamically preferred and that
the phase transition is generically second order. The helical order is fixed by wave-number $k$, or equivalently a pitch $p=2\pi/k$. We calculate $k(T)$ and find it
monotonically decreases down to a finite value as $T\to 0$. We find that the solutions have vanishing entropy density as $T\to 0$. In this limit they
approach smooth domain wall solutions, which we also construct, that interpolate between $AdS_5$ in the UV and
a new spatially homogeneous but non-isotropic ground state in the IR, of a type recently been discussed in \cite{Iizuka:2012iv}. 

There are other contexts in which spatially modulated black holes should exist but, in general, the construction will require solving non-linear partial differential equations. By contrast, a key point here is that the black holes for the helical $p$-wave superconductors can be obtained by solving ordinary differential equations since they are static and also have a Bianchi VII$_0$ symmetry.

\section{The D=5 model}
\label{sec:d=5model}
As in \cite{Donos:2011ff} we consider a $D=5$ model coupling a metric to a gauge field $A$ and a complex two-form $C$ with action
\begin{align}\label{eq:lag2f}
S=\int d^5 x\sqrt{-g}\big[R+12&-\tfrac{1}{4}F_{\mu\nu}F^{\mu\nu}-\tfrac{1}{4}C_{\mu\nu}\bar{C}^{\mu\nu}\nn
&+\tfrac{{i}}{24m}\epsilon^{\mu\nu\rho\sigma\delta}C_{\mu\nu}\bar{H}_{\rho\sigma\delta}\big]\,,
\end{align}
where a bar denotes complex conjugation and the field strengths, using a form notation, are given by
\begin{align}\label{ders}
F=dA,\qquad
H=dC+{i} e\,A\wedge C\,.
\end{align}

This simple class of models, specified by the parameters $m$ and $e$, is rather natural. 
The equations of motion admit a unit radius $AdS_5$ solution with $A=C=0$, which
is dual to some putative conformal field theory (CFT). The massless gauge-field $A$ is dual to a current in the CFT, corresponding to a global abelian symmetry,
with scaling dimension $\Delta=3$. 
The two-form $C$ satisfies a first-order equation of motion and is dual to a self-dual rank two tensor 
operator with $\Delta=2+|m|$. In particular, this charged operator has $l=1$ and thus provides an
order parameter for $p$-wave superconductivity. Such two-forms
are common in Kaluza-Klein reductions from D=10 or D=11 supergravity. For example, when $e=1/\sqrt{3}$, $m=1$ precisely 
this model can be obtained as a consistent Kaluza-Klein truncation of type IIB supergravity 
on $S^5$ and moreover the operator in $N=4$ SYM dual to $C$ is known \cite{Aprile:2010ge}.

We will study the CFT dual to the $AdS_5$ vacuum at finite temperature $T$ and chemical potential $\mu$ with respect to the global abelian symmetry by constructing electrically charged asymptotically $AdS_5$ black branes.

\section{Black hole solutions}
The ansatz for all the black hole solutions that we consider is given by
\begin{align}\label{eq:ansatz}
ds^{2}&=-g\,f^{2}\,dt^{2}+g^{-1}{dr^{2}}+h^{2}\,\omega_{1}^{2}+r^{2}\,\left(e^{2\alpha}\,\omega_{2}^{2}+e^{-2\alpha}\,\omega_{3}^{2}\right)\notag\\
C&=(i\, c_{1}\,dt+c_2 dr)\wedge\omega_{2}+c_{3}\,\omega_{1}\wedge\omega_{3}\,,\notag\\
A&=a\,dt\,,
\end{align}
where the one-forms $\omega_i$ are given by
\begin{align}\label{eq:one_forms}
&\omega_{1}=dx_{1}\,,\nn
&\omega_{2}=\cos\left(kx_{1}\right)\,dx_{2}-\sin\left(kx_{1}\right)\,dx_{3}\,,\nn
&\omega_{3}=\sin\left(kx_{1}\right)\,dx_{2}+\cos\left(kx_{1}\right)\,dx_{3}\,,
\end{align}
and $f$, $g$, $h$, $\alpha$, $c_{i}$ and $a$ are all functions of the radial coordinate, $r$, only and $k$ is a constant.
Observe that $k$ can be scaled out of the ansatz by scaling $h$ but has been included for later convenience.
The unit radius $AdS_5$ vacuum solution can be obtained by setting $g=r^2$, $f=1$, $h=r$, $\alpha=a=c_i=0$. 
 Notice that the ansatz \eqref{eq:ansatz}
 is static and in addition the constant $t$ and $r$ slices are spatially homogenous of, generically, Bianchi type VII$_0$.

Substituting this ansatz into the $D=5$ equations of motion we find that we can solve for $c_{1}$ and $c_{2}$:
\begin{align}
c_{1}=-\frac{e^{2\alpha}}{e^{4\alpha}k^{2}+m^{2}\,h^{2}}\,\left( e^{2\alpha}kea c_{3}+mhfgc_{3}^{\prime}\right)\,,\notag\\
c_{2}=\frac{1}{fg}\frac{e^{2\alpha}}{e^{4\alpha}k^{2}+m^{2}h^{2}}\,\left( meahc_{3}-e^{2\alpha}kfgc_{3}^{\prime}\right)\,.
\end{align}
The remaining equations can be obtained from a one-dimensional action obtained by substituting the ansatz into the action associated with 
\eqref{eq:lag2f}.

The equations of motion admit the electrically charged AdS-RN black brane solution with
$\alpha=c_i=0$, $f=1$, $h=r$ and
\begin{align}\label{adsrn}
g=r^{2}-\frac{r^{4}_{+}}{r^{2}}+\frac{\mu^{2}}{3}\,\left(\frac{r_{+}^{4}}{r^{4}}-\frac{r^{2}_{+}}{r^{2}} \right),
\,a=\mu\,\left(1-\frac{r_{+}^{2}}{r^{2}} \right).
\end{align}
This solution approaches the unit radius $AdS_5$ solution as $r\to\infty$.
The event horizon is located at $r=r_+$ and the temperature is $T=(6r_+^2-\mu^2)/6\pi r_+$. This solution is dual to the CFT at finite chemical potential $\mu$ and
high temperatures. Clearly this phase is spatially homogeneous and isotropic. It was shown in \cite{Donos:2011ff}
that below a critical temperature, depending on the parameters $m,e$, and the scale set by $\mu$,
this black hole is unstable to the formation of black holes that are dual to $p$-wave superconductors with helical order. 

Let us first discuss the boundary conditions to be imposed for the new black brane solutions. Regularity at the horizon demands that
$g(r_+)=a(r_+)=0$. We then find the solution at the horizon is specified by six parameters: $r_+$,  
$f(r_+)$, $h(r_+)$, $\alpha(r_+)$, $a'(r_+)$ and $c_3(r_+)$.
As $r\to \infty$ we approach $AdS_5$ with asymptotic expansion
\begin{align}\label{uvexp}
&g=r^{2}\,\left(1-{M}{r^{-4}}+\cdots \right),\quad
f=f_{0}\left(1-{c_{h}}{r^{-4}}+\cdots\notag\right),\\
&h=r\,\left(1+{c_{h}}{r^{-4}}+\cdots \right),\quad
\alpha={c_{\alpha}}{r^{-4}}+\cdots,\notag\\
&a=f_{0}\,\left(\mu+{q}{r^{-2}}+\cdots\right),\quad
c_{3}={c_{v}}{r^{-\left|m\right|}}+\cdots,
\end{align}
which is specified by eight
parameters $M, f_0,c_h,c_\alpha,\mu,q,c_v$ and $k$. 
A number of comments are in order. Firstly, the fact that $h\sim r$ implies that the wave-number $k$ can no longer be scaled away.
Secondly, the fall-off of $c_3$ is chosen so that the charged operator dual to the two-form 
$C$ has no deformation but can acquire, {\it spontaneously}, an expectation value proportional to $c_v$ and spatially modulated in the $x_1$ direction with period $2\pi/k$. The holographic interpretation of the other UV parameters will be given below.
Thirdly, there are two scaling symmetries of the differential equations which allow us to set $\mu=f_0=1$, and we will do so
later (it is helpful to have them to discuss the thermodynamics). 
Finally, we have four second order differential equations for $h,\alpha,a,c_3$ and 
two first order equations for $g,f$ and hence a solution is specified by ten integrations constants. On the other hand we have fourteen parameters in the boundary conditions minus two for the scaling symmetries. We thus expect a two parameter family of black hole solutions
that can be specified by temperature $T$ and wave number $k$.

\subsection{Action and Thermodynamics}
We analytically continue by setting $t=-i\tau$ and defining $I=-iS$. The total action, including
relevant boundary-terms, is given by
\begin{align}\label{counter term}
I_{Tot}=I+\int d\tau d^3x\sqrt{g_\infty}[-2K+6+\dots]\,,
\end{align}
with $g_\infty=\lim_{r\to\infty}g^{1/2}fh r^2$ and $K$ is the trace of the extrinsic curvature of the boundary at $r\to\infty$. In \eqref{counter term} the ellipses refer to terms that will not contribute to the
class of solutions that we are considering. The period of our Euclidean time is taken to be $\Delta \tau$ and the temperature is then given by $T=(f_0\Delta \tau)^{-1}$. Regularity at $r=r_+$ implies that $T=\frac{f}{f_0}g'(4\pi)^{-1}|_{r=r_+}$.
We next define the thermodynamic potential $W=T[I_{Tot}]_{OS}\equiv w \mathrm{vol}_3$, where $[I_{Tot}]_{OS}$ is the on-shell action.
Following the calculation in \cite{Gauntlett:2009bh} we obtain the two equivalent 
expressions
\begin{align}\label{smarr}
w=-M=\varepsilon+2\mu q-Ts\,,
\end{align}
where we defined the entropy density $s=4\pi r^{2}h_{+}$, $\varepsilon=3M+8c_{h}$ and have set $f_0=1$.
By calculating the on-shell variation of 
$I_{Tot}$, including variations of $f_0$, we deduce that $w=w(T,\mu)$ and the first law
\begin{align}
\delta w=-s\delta T+2q\delta\mu\,.
\end{align}
Using \eqref{smarr} we also have $\delta\epsilon=T\delta s-2\mu\delta q$. The identification of $\varepsilon$
with energy density is confirmed by computing the boundary stress-energy tensor \cite{Balasubramanian:1999re}, again with $f_0=1$,
\begin{align}
&T_{tt}=3M+8c_{h}\,,\quad
T_{x_{1}x_{1}}=M+8c_{h}\,,\notag\\
&T_{x_{2}x_{2}}=M +8c_{\alpha}\,\cos\left(2kx_{1} \right) \,,\notag\\
&T_{x_{3}x_{3}}=M -8c_{\alpha}\,\cos\left(2kx_{1} \right) ,\notag\\
&T_{x_{2}x_{3}}=-\,8\,c_{\alpha}\,\sin\left(2kx_{1}\right)\,.
\end{align}
Observe that when $c_\alpha\ne 0$ the pressures in the $x_2$, $x_3$ plane are spatially modulated as one moves along the $x_1$ direction as one expects
for helical order. Defining the average hydrostatic pressure, $\bar p$, as minus the average of the trace of the spatial components we get $\bar p=M+8c_h/3$.
The thermodynamically preferred black hole solutions that we construct will have $c_\alpha\ne 0$ and $c_h=0$. Using
\eqref{smarr} we conclude that this class satisfy the thermodynamic relation 
$
\varepsilon+\bar p=Ts-2\mu q$.

\subsection{Helical superconducting black holes}
The AdS-RN black brane solution \eqref{adsrn} is unstable when $e^2>m^2/2$ \cite{Donos:2011ff}. We will now consider the specific model with $m=1.7$, $e=1.88$ (for reasons we explain below) and set $\mu=f_0=1$.
For high temperatures we only find the AdS-RN black hole solution. The first new black hole solution appears at $T_c\approx 0.0265$
and for $k=k_c\approx 0.550$. Holding this value of $k$ fixed we construct these black hole solutions, numerically, 
all the way down to very low temperatures.
Below $T_c$, as expected from the linearised analysis of \cite{Donos:2011ff}, there is a continuum of black hole solutions that appear with different values of $k$. By again holding $k$ fixed we can construct each of these black holes too, down to low-temperatures. 

In fig. 1 we have summarised this new two-parameter family of solutions as well as displaying their free-energy $w$.
\begin{figure}[t!]
\includegraphics[width=0.4\textwidth]{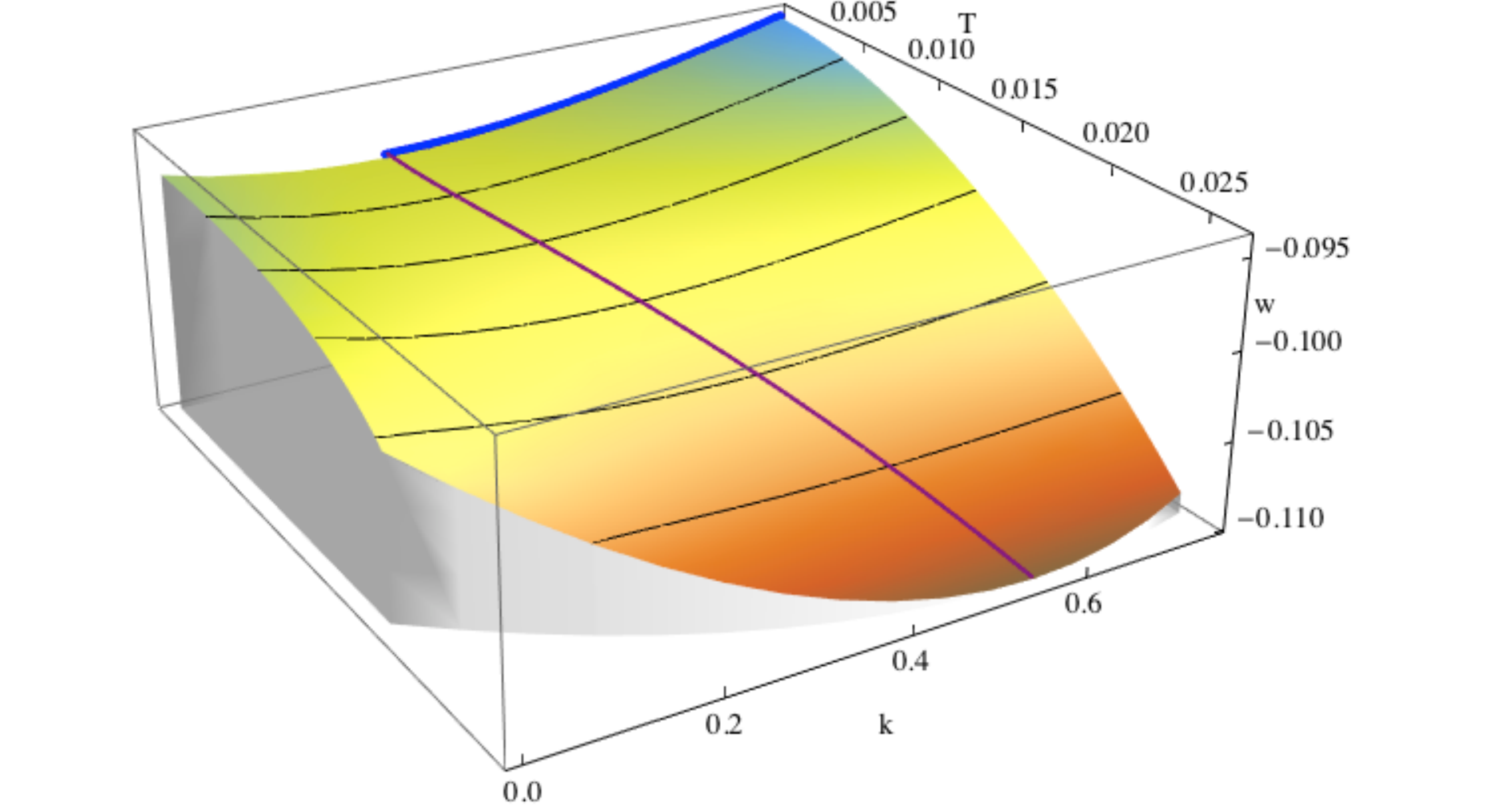}
\caption{The two-parameter family of helical superconducting black holes and their free-energy $w$.
The red line denotes the thermodynamically preferred locus. The blue line is the free-energy of some domain wall solutions. The black lines are lines of constant $T$.
\label{fig:1}}
\end{figure}
All of these solutions have smaller free energy than the AdS-RN black brane solutions at the same temperature.
At a given temperature $T<T_c$ there is a one parameter family of black hole solutions specified by $k$ and the one with the smallest 
free energy is depicted by a point on the red line in fig. 1. Thus the one-parameter family of solutions specified by the red line
characterises the thermodynamically preferred solutions. Notice that we have a second order phase transition at $T=T_c$, $k=k_c$ and
that as the temperature is lowered, the system smoothly moves between
black hole solutions with different values of $k$, all the way down to very low temperature where $k\equiv k_0\approx 0.256$. In particular, the 
$T=0$ ground state remains spatially modulated.

Interestingly, while the general two-parameter family of solutions have $c_h\ne 0$, the solutions on the red line have
(up to numerical accuracy) $c_h=0$.
In fig. 2 we have plotted, for the red line of solutions, the behaviour of $c_v$ and wave-number $k$, which, together, characterise 
the helical superconducting order, versus $T$. Near $T_c$ we find the mean field behaviour
$c_{v}\approx1.7\times 10^5 T_c^{3.7}\left(1-T/T_{c} \right)^{1/2}$.
\begin{figure}[t!]
\includegraphics[width=0.22\textwidth]{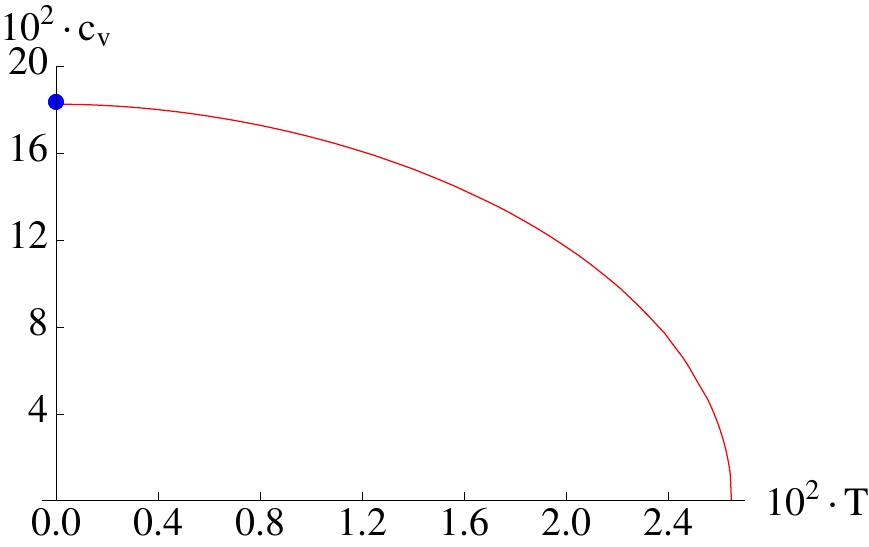}
\hskip1em \includegraphics[width=0.22\textwidth]{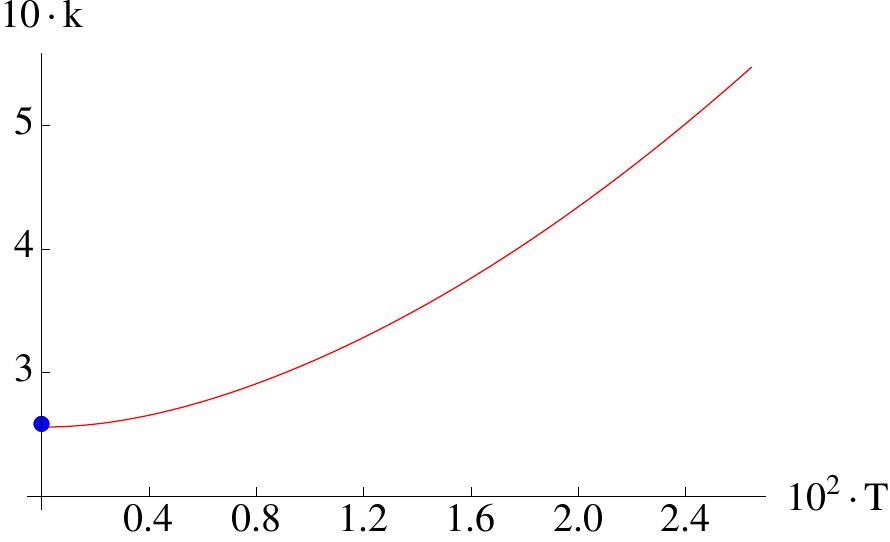}
\caption{Plots of $c_v$ and wave-number $k$, which together fix the helical superconducting order, versus T for the thermodynamically preferred black hole solutions on the red line in fig. 1. The blue dots depict the quantities for the domain wall solutions. Note the scaled axes.
\label{fig:2}}
\end{figure}

\subsection{New ground states at $T=0$}
We are particularly 
interested in the $T= 0$ limit of the thermodynamically preferred black hole solutions (the red line in fig. 1), 
which has $k=k_0\approx 0.256$. It is 
helpful to first consider the $T=0$ limit of the whole class of black holes for general values of $k$. 
We find that they all approach a smooth domain wall solution that interpolates between $AdS_5$ in the UV and a new fixed point in the IR with a scaling symmetry of a type that is very similar to those of \cite{Iizuka:2012iv}. Indeed we checked this explicitly for the range $0.253\le k\le 0.75$ (going to smaller values
of $k$ becomes increasingly difficult numerically).

To obtain this new fixed point solution, in \eqref{eq:ansatz} we put
\begin{align}\label{lif}
&g=L\,r^{2},\quad f=\bar f_0r^{z-1},\quad h=k h_{0}\,,\quad  \alpha=\alpha_{0}\,,\nn
&a=a_{0}r^{z},\quad
c_{3}=k c_{0}\,r\,,
\end{align}
where $L,h_0,\alpha_0,a_0,c_0$ and $z$ are all constant. By scaling $t$ and $x_1$ we can set $\bar f_0=k=1$.
Notice that this ansatz corresponds to a solution invariant under the anisotropic scaling
$r\rightarrow\lambda^{-1}r$, $t\rightarrow \lambda^{z}t$, $x_{2,3}\rightarrow\lambda x_{2,3}$ and $x_1\to x_1$.
After substituting into the equations of motion we obtain a system of algebraic equations which can be solved. 
For the specific case of $m=1.7$, $e=1.88$ we find 
\begin{align}
&z\approx1.65 \ldots,\qquad L\approx 0.995\ldots,\qquad h_{0}\approx0.993\ldots,\nn
&\alpha_{0}\approx-0.380\ldots,\quad a_{0}\approx0.265\ldots, \quad 
 c_{0}\approx 3.69\ldots
\end{align}

We next construct domain wall solutions that interpolate between this fixed point in the IR and $AdS_5$ in the UV\footnote{
Domain walls were constructed for a different D=5 model in \cite{Iizuka:2012iv}. However, unlike our solutions, the breaking of the spatial Euclidean symmetry to Bianchi VII$_0$ symmetry is not spontaneous.}.
In the UV we continue to demand the expansion given by \eqref{uvexp}. To obtain the IR expansion, we first consider
perturbations about \eqref{lif} of the form
\begin{align}
&g=r^{2}\,\left(L+\lambda w_{1} r^{\delta} \right),\qquad f=\bar f_{0}\,r^{z-1}\left(1+\lambda w_{2}r^{\delta} \right)\,,\notag\\
&h=k\left(h_{0}+\lambda w_{3}r^{\delta}\right),\qquad
\alpha=\alpha_{0}+\lambda w_{4}r^{\delta}\,,\notag\\
&a=f_{0}a_{0}r^{z}\,\left(1+\lambda w_{5}r^{\delta} \right),\,\,
c_{3}=kc_{0}r\,\left(1+\lambda w_{6}r^{\delta} \right).
\end{align}
After expanding the equations of motion at first order in $\lambda$ we obtain a homogeneous linear system of equations $\mathbf{E}\cdot\mathbf{w}=0$ where $\mathbf{E}$ is a $6\times 6$ matrix  that depends on $\delta$.
Demanding non-trivial solutions for $\mathbf{w}$ we determine the values of $\delta$ by solving the polynomial equation $\left| \mathbf{E}\right|=0$.
For the special case $m=1.7$, $e=1.88$ the modes with non-negative real parts have $\delta_{0}=0$, $\delta_{1}\approx 0.394$, $\delta_{2}\approx 0.826$, $\delta_{3}\approx 0.847$ and $\delta_{4}\approx 2.289$ \footnote{Obtaining real scaling dimensions was the main reason for
focussing on $m=1.7$, $e=1.88$. For many
values of $m,e$ there will be complex scaling dimensions analogous to \cite{Gubser:2009cg}. An example is $m=2$, $e=2$ and for this case we have also constructed all of the black holes and domain wall solutions, with significantly easier numerics, finding a very similar story with the $T=0$ ground state matching onto a scaling solution with $z\approx1.99$.}.
The IR expansion is then specified by a constant for each of these modes. Notice that 
the mode with $\delta_{0}=0$ corresponds to the constant $\bar f_{0}$. This leads to five (real) parameters in the IR. With the eight parameters
mentioned for the UV discussed earlier, we now deduce that the domain wall solutions will be specified by a single parameter
which we can take to be the wave-number $k$.

We have constructed these solutions for the range $0.253\le k\le 0.75$, which required utilising high precision numerics.
As noted above we find that the $T=0$ limit of the black hole solutions approach these domain wall solutions.
For example, in fig. 1 the blue line depicts the free energy of the domain wall solutions for various values of $k$, showing
precise agreement with the $T=0$ limit of the black holes. Similarly, in fig. 2, for the domain wall with $k=k_0\approx 0.256$
we have shown the UV values of $c_v$ and $k$ with a blue dot and again we see precise agreement with the corresponding black hole 
solution.

\section{Final Comments}
Let us summarise the main physical results. The $d=4$ CFT, dual to the $AdS_5$ solution of our model \eqref{eq:lag2f}, 
held at finite chemical potential undergoes a second order phase transition at a critical temperature $T_c$. The new phase is a helical superconducting phase that spontaneously breaks both the global abelian symmetry and the three-dimensional spatial Euclidean symmetry down to Bianchi VII$_0$ symmetry. At $T_c$ the spatial modulation is fixed by a
wave number $k_c$ and as the temperature is cooled, the wave-number monotonically decreases. 
The $T=0$ ground state of the system maintains
the helical order with non-vanishing wave number and has an emergent scaling symmetry in the far IR. These homogeneous, non-isotropic ground states at $T=0$  
are holographically described by smooth domain wall solutions. 
It is natural to next use the AdS/CFT correspondence to analyse transport. Based on the rich
optical properties of other helical orders (such as the chiral nematic phase of liquid crystals, recently discussed in \cite{Gibbons:2011im}) we expect interesting results.

We expect analogous solutions for the helical superconductors of the $D=5$ models with $SU(2)\times U(1)$ gauge fields
studied in \cite{Donos:2011ff}. In particular we have checked that the (top-down) Romans' theory admits a
Bianchi VII$_0$ ground state solution with scaling exponent $z\approx 3.98$ which we conjecture to be the
IR limit of a $T=0$ domain wall solution. 
More generally we now anticipate many other constructions of black holes in $D=5,4$ that are dual to other spatially modulated phases, both superconducting and otherwise.
\subsection*{Acknowledgements}
We thank S. Gubser, D. Waldram and T. Wiseman for helpful discussions.
AD is supported by an EPSRC Fellowship.
JPG is supported by a Royal Society Wolfson Award.

\end{document}